\documentclass[12pt,onecolumn]{IEEEtran}
\usepackage{amssymb}
\usepackage{amsmath}
\usepackage{amsthm}
\usepackage{amsfonts}
\usepackage{amsthm}
\usepackage{amsmath}
\usepackage{amsfonts}
\usepackage{amssymb}
\usepackage{graphicx}
\usepackage{hyperref}
\usepackage{mathtools}
\usepackage{verbatim}
\usepackage{subcaption}
\usepackage{float}
\usepackage{eucal}
\usepackage{balance}
\usepackage{bm}
\usepackage{amssymb}
\usepackage{amsmath}
\usepackage{amsthm}
\usepackage{amsfonts}
\usepackage{hyperref}
\usepackage{mathtools}
\usepackage{verbatim}
\usepackage{subcaption}
\usepackage{txfonts}
\usepackage{xfrac}
\usepackage[noadjust]{cite}
\usepackage{amsmath,amssymb,amsfonts}
\usepackage{algorithmic}
\usepackage{graphicx}
\usepackage{textcomp}
\usepackage{float}
\usepackage{dblfloatfix}
\usepackage{authblk}
\usepackage{bm}
\allowdisplaybreaks

\newtheorem{theorem}{Theorem}

\newtheorem{definition}[theorem]{Definition}

\newcommand{\E}{\mathbb{E} \,\,}

\title{Low SNR Capacity of Keyhole MIMO Channel\\ in Nakagami-$m$ Fading With Full CSI}

\author{
\IEEEauthorblockN{Kamal Singh\IEEEauthorrefmark{1}, Chandradeep Singh\IEEEauthorrefmark{2}, Kuang-Hao Liu\IEEEauthorrefmark{3}}
\thanks{

\IEEEauthorrefmark{1}{Department of Electrical Engineering, Shiv Nadar University, Delhi NCR, India-201314 (e-mail: kamal.singh@snu.edu.in).}

\IEEEauthorrefmark{2}{Department of Electrical Engineering, National Cheng Kung University, Tainan 701, Taiwan (e-mail: chandradeep.chd@gmail.com).}

\IEEEauthorrefmark{3}{Department of Electrical Engineering, National Tsing Hua University, Hsinchu 30013, Taiwan (email: khliu@ee.nthu.edu.tw).}}
\vspace*{-0.75cm}
}

\begin{document}
\maketitle
\begin{abstract}
In this paper, we obtain asymptotic expressions for the ergodic capacity of the keyhole multiple-input multiple-output (MIMO) channel at low signal-to-noise ratio (SNR) in independent and identically distributed Nakagami-$m$ fading conditions with perfect channel state information at the transmitter and receiver. We show that the low-SNR capacity of this keyhole MIMO channel scales proportionally as $\frac{\textrm{SNR}}{4} \log^2 \left(1/{\textrm{SNR}}\right)$. Our main contribution is to identify a surprising result that the low-SNR capacity of the MIMO fading channel increases in the presence of keyhole degenerate condition, which is in direct contrast to the well-known MIMO capacity degradation at high SNR under keyhole conditions. To explain why rank-deficient keyhole fading channel outperforms the full-rank MIMO fading channel at sufficiently low-SNR, we remark that the rank of the MIMO channel matrix has no impact in the low-SNR regime and that the double-faded (or double-scattering) nature of the keyhole MIMO channel creates more opportunistic communications at low-SNR when compared with pure MIMO fading channel which leads to increased capacity. Finally, we also show that a simple one-bit channel information based on-off power control achieves this low-SNR capacity; surprisingly, this power adaptation is robust against both moderate and severe fading  for a wide range of low SNR values. These results also hold for the keyhole MIMO Rayleigh channel as a special case.
\end{abstract}
\vspace{-0.2cm}
\section{Introduction}
\label{sec:introduction}
\begin{figure*}[!t]
\centering
\includegraphics[scale = 1.65]{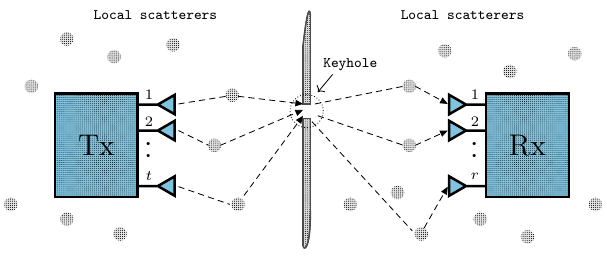}
\vspace*{1em}
\caption{Keyhole MIMO channel with $t$ transmit and $r$ receive antennas.}
\label{fig:keyhole}
\end{figure*}

\IEEEPARstart{T}{he} multiple-antenna systems (a.k.a. MIMO systems) generally provide manifold increase in the channel capacity over single-antenna systems subject to the presence of rich scattering wireless channel and sufficient antenna spacing at both transmitter and receiver~\cite{telatar_mimo},~\cite{foschini}. Specifically, it is shown that the capacity of a MIMO system with fixed transmit power budget and subjected to independent and identically distributed (IID) Rayleigh fadings scales linearly with the minimum of the number of transmit and receive antennas in the moderate-to-high SNR regime~\cite{telatar_mimo}. This is due to the full-rank realization of MIMO fading channel made possible in rich scattering conditions. However, in realistic propagation environments, this full-rank realization might not be always possible if degenerate channel conditions exist. For example, the presence of spatial fading correlation and/or \emph{keyhole effect} can degenerate the full-rank property of MIMO channel, thereby severely degrading the spectral efficiency of the MIMO systems~\cite{chua}-\cite{shinlee}. The spatial fading correlation arises due to insufficient spatial seperation among antenna elements and/or lack of scattering: capacity is low for highly correlated fadings and it is high for very low fading correlation between antenna pairs~\cite{chua}-\cite{molisch_corr}. A keyhole scenario arises when the radio wave propagation from the transmitter towards the receiver is possible only via passing through a small keyhole (see Fig.~\ref{fig:keyhole}). The presence of keyhole, regardless of correlation, reduces the spatial multiplexing gain (or rank) of MIMO fading channel to unity, which is termed as `keyhole effect' in the literature (see~\cite{gesbert},~\cite{chizhik2} for details). Hence, from a capacity viewpoint, the MIMO channel model under keyhole effect is generally regarded in the literature as as the worst-case MIMO propagation. We remind the reader that although the keyhole effect in MIMO systems is not prevalent in most practical fading environments, the phenomenon is theoretically significant due to its unique position to represent the worst-case MIMO propagation. An illustration of a realistic keyhole condition in an outdoor propagation environment is shown in Fig. ~\ref{fig:keyhole} that may arise due to propagation in two local rich scattering environments connected by a corridor/tunnel or diffractions over an edge between two rich scattering environments. This keyhole degenerate condition in MIMO fading environments has been theoretically predicted by Gesbert \emph{et al.}~\cite{gesbert} and Chizhik \emph{et al.}~\cite{chizhik2}, and extensive measurements in a controlled indoor environment by
Almers \emph{et al.}~\cite{almers0},~\cite{almers} fully validated the keyhole effect. Besides its relevance for the aforementioned MIMO propagation scenarios, the keyhole channel can also be used to model the relay channel in the amplify-and-forward mode in some practically relevant scenarios as explained in detail in~\cite{levin}.

In this paper, we will consider the keyhole MIMO channel in Nakagami-$m$ fading conditions under the assumption that the instantaneous keyhole channel state is estimated perfectly at the receiver and this channel knowledge (exact) is made available to the transmitter by providing an error-free delayless feedback path. This widely used assumption of full channel knowledge allows transmitter to adapt its signal characteristics according to the channel state at that instant. Most previous research that deals with the capacity analysis of keyhole MIMO channels under various assumptions on channel knowledge at the transmitter includes the following. The capacity of keyhole MIMO channels with and without perfect channel knowledge at the transmitter in correlated Rayleigh fading are investigated in~\cite{aissa} and~\cite{shinlee} respectively; particularly, for the special case of independent and identically distributed (IID) Rayleigh fading, exact closed-form capacity expressions are derived. The keyhole MIMO channel capacity at low-SNR for correlated Rayleigh fading is analyzed in~\cite{jin1} under the assumption of varying quantized levels of channel knowledge at the transmitter. The capacity of the more general Rayleigh Product MIMO channel (keyhole MIMO channel becomes a special case, see~\cite{gesbert} for definition) with transmit beamforming is investigated thoroughly in~\cite{jin}, and its behaviour at low-SNR in the presence of a co-channel interferer and without channel knowledge at the transmitter is studied in~\cite{ratnarajah}. Although significant research efforts have been carried out to analyze the capacity of keyhole MIMO channels in Rayleigh fading, the existing literature on the capacity of keyhole MIMO channels in the more general and empirically-fit Nakagami-$m$ fading conditions is rather limited and the only capacity results of which we are aware are presented in~\cite{muller}, derived for independent Nakagami-$m$ fadings and without channel knowledge at the transmitter. The conspicuous lack of closed-form capacity expressions for the case of full channel knowledge at the transmitter can be attributed to analytical difficulties in simplifying the capacity (integral) expression that involves complicated Nakagami-$m$ keyhole channel's distribution function. Nevertheless, as an alternate approach, asymptotic capacity analysis in the extreme SNR regimes can provide useful insights as such a characterization generally reveals dependence applicable for moderate conditions.

While the adverse effect of the degenerate keyhole condition on the MIMO fading channel capacity in the high-SNR regime is well explained in terms of the `collapse' of the \emph{spatial degrees of freedom gain} to unity, there is limited knowledge available in the existing literature about the role of the keyhole effect on the MIMO fading channel capacity in the low-SNR regime where spatial degrees of freedom gain has little or no impact. Motivated by the importance of understanding the capacity of keyhole MIMO channels, in this paper, we will derive asymptotic capacity expressions of the keyhole MIMO channel at low-SNR in IID Nakagami-$m$ fading conditions under full channel knowledge assumption at the transmitter and receiver. The low-SNR framework encompasses all relevant communication scenarios operating in severe fading like in wireless cellular networks in some specific cases~\cite{lozano}, in wireless sensor networks operating with a very low power budget~\cite{karaki} or more generally in any communications with limited power and fixed bandwidth resources such that the power per degree of freedom is very low. This is especially true for wideband communication systems with limited power and very large system degrees of freedom such that the power available per degree of freedom is extremely low~\cite{verdu}. Note that a keyhole channel with only single degree of freedom fades twice as often as a normal IID channel and thus, it is more vulnerable to weak SNR conditions. Nevertheless, it is encouraging to note that in the low-SNR regime, the capacity for a wide class of fading channels is significantly larger with channel knowledge at the transmitter than  without it; varying transmit power as a function of the channel state is highly beneficial at low SNRs~\cite{tsebook}. Our specific contributions are summarized as follows:\vspace{0.35em}
\begin{itemize}
\item For the keyhole MIMO channel in Nakagami-$m$ fading and with full channel knowledge at the transmitter and receiver, we derive two asymptotic low-SNR capacity expressions: one in terms of the Lambert $W$ function and the second in terms of the Log function.\\[-0.65em]
\item At asymptotically low-SNR, the keyhole MIMO channel capacity in Nakagami-$m$ fading is shown to scale proportionally as $\frac{\textrm{SNR}}{4} \log^2 \left(\frac{1}{\textrm{SNR}}\right)$. Further, the higher the severity of the Nakagami-$m$ fading condition, the larger the low-SNR capacity of the keyhole MIMO channel.\\[-0.65em]
\item A simple 1-bit channel information based on-off transmission is shown to be capacity achieving at low-SNR. Surprisingly, this power adaptation is robust against both moderate and severe Nakagami-$m$ fadings for a wide range of low-SNR values.\\[-0.65em]
\item Most importantly, based on these asymptotic capacity results, we derive a very surprising conclusion that in the low-SNR regime, MIMO fading channel capacity increases in the presence of degenerate keyhole condition.\\[-0.85em]
\end{itemize}

The remainder of this paper is organized as follows. In Section~\ref{sec:intro}, we outline the keyhole MIMO channel and system model. In Section~\ref{sec:three}, we perform  low-SNR capacity analysis of the keyhole MIMO channel in Nakagami-$m$ fading with channel knowledge assumption at the transmitter and receiver, followed by numerical results illustrating the accuracy of the proposed asymptotic low-SNR capacity formulae. Section~\ref{sec:3B} describes the asymptotically (at low-SNR) optimal on-off transmission policy for the keyhole channel. In Section~\ref{sec:four}, we analyze the impact of keyhole degeneracy on the capacity of MIMO fading channel in the low-SNR regime. Section~\ref{sec:conc} offers some concluding remarks.

A note on notation: We will use boldface letters to denote random variable, random vector and random matrix and their corresponding realizations are denoted by the same letters but without boldface.  $\mathbb{E}_{\textbf{X}} [\cdot]$ denotes expectation and $f_{\textbf{X}} (\cdot)$ the corresponding PDF with respect to the random variable $\textbf{X}$. $\mathbf{I}$ denotes the identity matrix, $\mathrm{det}(A)$ and $\mathrm{rank}(A)$ denote the determinant and rank of the matrix $A$. $\log(\cdot)$ denotes the natural logarithm operation and $\mathrm{max}\{0,x\}$ by $x^{+}$. We will use the abbreviation $\{x_i \}_{i=1}^{K}$ for the vector $[x_1,\dots,x_K]$.

\section{System and Channel Model}\label{sec:intro}
We consider a double-scattering keyhole MIMO channel as shown in Fig.~\ref{fig:keyhole} with perfect channel knowledge at both sides and subjected to flat independent Nakagami-$m$ fadings. With $t$ transmit and $r$ receive antennas, the received signal vector is described as
\begin{align}\label{eq:sys_model}
\textbf{y} = \textbf{H} \textbf{x} + \textbf{w}
\end{align}
where $\textbf{H} \in \mathbb{C}^{r \times t}$ is the channel matrix generated by an ergodic stationary process, $\textbf{x} \in \mathbb{C}^t$ is the channel input, $\textbf{y} \in \mathbb{C}^r$ is the channel output and $\textbf{w} \in \mathbb{C}^r$ is zero-mean complex Gaussian noise with independent, equal variance real and imaginary parts, and $\mathbb{E}_{\textbf{w}}\, [\textbf{w}\textbf{w}^{\dagger}] = \mathbf{I}_r$. The input $\textbf{x}$ is subjected to the average power constraint $P_{\mathrm{avg}}$, i.e., $\mathbb{E}_{\textbf{x}}\, [\textbf{x}^\dagger \textbf{x}] = P_{\mathrm{avg}}$.

The transmitted signal can propagate to the receiver side only via the keyhole connecting the two rich scattering environments. We assume that the keyhole reradiates the captured energy like an ideal scatterer, see Fig.~\ref{fig:keyhole}. The keyhole MIMO channel is then described as
\begin{align}\label{eq:keyholeH}
\textbf{H} := \textbf{h}_r \textbf{h}_t^{T}
\end{align}
where $\textbf{h}_r := \{\boldsymbol{\beta}_i e^{j \boldsymbol{\phi}_i}\}_{i=1}^{r}$ and $\textbf{h}_t := \{\boldsymbol{\alpha}_l e^{j \boldsymbol{\psi}_l}\}_{l=1}^{t}$ denote the channel vectors from the keyhole-to-receiver and transmitter-to-keyhole respectively.
In our channel model, we assume that all the entries of the channel vector $\textbf{h}_t$ are IID distributed; the probability density function (PDF) of the magnitude is according to the Nakagami-$m$ fading distribution with parameters $(m_t,\Omega_t)$~\cite{Nakagami_dist} and the phase is uniformly distributed in $[0,2\pi)$. Thus, $\forall l$,
\begin{align}\label{eq:dist:naka1}
f_{\boldsymbol{\alpha}_{l}} (\alpha) &= \dfrac{2}{\Gamma(m_t)}\left(\dfrac{m_t}{\Omega_t}\right)^{m_t} \alpha^{2m_t - 1} e^{-\tfrac{m_t}{\Omega_t}\alpha^2},\,\,\alpha \geq 0
\end{align}
where $m_t \geq 1/2$ and $\Omega_t > 0$ are the shape and scale parameters of the Nakagami-$m$
distribution respectively, and $\Gamma(\cdot)$ is the Gamma function~\cite{tableofintegrals}.
Likewise, we make a reasonable IID Nakagami-$m$ fading assumption on all the magnitude entries $\boldsymbol{\beta}_{i},\,i = 1,\dots,\,r$ in the channel vector $\textbf{h}_r$ with $m_r$ and $\Omega_r$ parameters as follows:
\begin{align}\label{eq:dist:naka2}
f_{\boldsymbol{\beta}_{i}} (\beta) &= \dfrac{2}{\Gamma(m_r)}\left(\dfrac{m_r}{\Omega_r}\right)^{m_r} \beta^{2m_r - 1} e^{-\tfrac{m_r}{\Omega_r}\beta^2},\,\,\beta \geq 0.
\end{align}
Note that $\Omega_t = \mathbb{E}_{\boldsymbol{\alpha}} [\alpha^2]$ and $\Omega_r = \mathbb{E}_{\boldsymbol{\beta}} [\beta^2]$. Further, the shape parameter ($m_t$ and $m_r$) controls the depth or severity of the envelope attenuation. The Rayleigh fading distribution is a special case when $m_r = 1$ and $m_t = 1$; values lesser or greater compared to one correspond to fading more severe or less severe than Rayleigh fading~\cite{Nakagami_dist}.
The $r \times t$ keyhole MIMO channel matrix $\textbf{H}$ is, thus, given by
\begin{align}
\textbf{H}  =    \begin{bmatrix}
\alpha_1 \beta_1 e^{j(\phi_1 + \psi_1)}   &  \alpha_2 \beta_1 e^{j(\phi_2 + \psi_1)}   & \dots  &   \alpha_t \beta_1 e^{j(\phi_t + \psi_1)}\\[0.5em]
\alpha_1 \beta_2 e^{j(\phi_1 + \psi_2)}  &  \alpha_2 \beta_2 e^{j(\phi_2 + \psi_2)}   &  \dots  &  \alpha_t \beta_2 e^{j(\phi_t + \psi_2)}\\
\dot{•}   &  \dot{•}   &\hspace{-0.65em}  \dot{•}  &  \dot{•}\\
\dot{•}   &  \dot{•}   &   \dot{•} &   \dot{•}\\
\dot{•}   &  \dot{•}   & \hspace{0.65em} \dot{•} &   \dot{•}\\
\alpha_1 \beta_r e^{j(\phi_1 + \psi_r)}   &  \alpha_2 \beta_r e^{j(\phi_2 + \psi_r)}   &  \dots &   \alpha_t \beta_r e^{j(\phi_t + \psi_r)}
\end{bmatrix}
\end{align}
with rank unity, i.e., $\mathrm{rank}(\textbf{H}) = 1$ as all the column vectors are linearly dependent. Also, all the entries in $\textbf{H}$ above are uncorrelated as all the channel fading coefficients $\{\alpha_l\}_{l=1}^{t}$ and $\{\beta_i \}_{i=1}^{r}$ are statistically independent. Hence, the capacity of the keyhole MIMO channel $\textbf{H}$ with perfect channel knowledge at the transmitter and receiver gets simplified as follows~\cite[Chapter~8]{tsebook}
\begin{align}
C &= \mathbb{E}_{\textbf{H}} [\log \det (\bm{I}_{r} + \textbf{H} P(\textbf{H})\textbf{H}^{\dagger})]\\
  &= \mathbb{E}_{\textbf{H}} [\log \det (\bm{I}_{r} + \textbf{h}_r \textbf{h}_t^{T} P(\textbf{h}_r \textbf{h}_t^{T})(\textbf{h}_r \textbf{h}_t^{T})^{\dagger})]\\
  &= \mathbb{E}_{\boldsymbol{\lambda}} [\log(1 +\boldsymbol{\lambda} P(\boldsymbol{\lambda}))]\label{eq:cap:csit_do}
\end{align}
where $\boldsymbol{\lambda}:= \lVert \textbf{h}_t \rVert^2  \lVert \textbf{h}_r \rVert^2$ is the effective channel gain and $P(\boldsymbol{\lambda})$ is, in effect, the \emph{optimal} power allocation scheme obeying the average power budget as
\begin{align}\label{eq:power_const_a}
\mathbb{E}_{\boldsymbol{\lambda}} [P(\boldsymbol{\lambda})] = P_{\mathrm{avg}}.
\end{align}
Notice that $\lVert \textbf{h}_t \rVert^2 = \sum_{l = 1}^{t} \alpha_l^2$ and $\lVert \textbf{h}_r \rVert^2 = \sum_{i = 1}^{r} \beta_i^2$. The squared Nakagami-$m$ variables $\alpha_l^2$ and $\beta_i^2$ follow Gamma distribution,\footnote{We use the notation $Z \sim  \Upsilon(\Omega,m)$ to denote the Gamma distribution as $
 f_Z(z) = \frac{1}{\Gamma(m) \Omega^m} z^{m-1} e^{-z/\Omega},\,\,\,\,z \, \geq \, 0
 $ where $m > 0$ and $\Omega > 0$ are the shape and scale parameters respectively, see~\cite{Zwillinger} for more details.} i.e., $\alpha_l^2 \sim \Upsilon(\Omega_t/m_t,m_t),\,\forall l$ and $\beta_i^2 \sim \Upsilon(\Omega_r/m_r,m_r),\,
\forall i$. The sums $\sum_{l = 1}^{t} \alpha_l^2$ and $\sum_{i = 1}^{r} \beta_i^2$ of IID Gamma variables
are also Gamma distributed; $\sum_{l = 1}^{t} \alpha_l^2 \sim \Upsilon(\Omega_t/m_t,t m_t)$ and $\sum_{i = 1}^{r}
\beta_i^2 \sim \Upsilon(\Omega_r/m_r, r m_r)$. Finally, the PDF of the effective fading gain $\boldsymbol{\lambda}$, which is equal to the product of the mutually independent non-negative random variables $\lVert \textbf{h}_t \rVert^2 := \sum_{l = 1}^{t} \alpha_l^2$ and $\lVert \textbf{h}_t \rVert^2 := \sum_{i = 1}^{r} \beta_i^2$, is obtained as
\begin{align}
f_{\boldsymbol{\lambda}} (\lambda) &= \int_{0}^{\infty} f_{\lVert\bm{h}_t\rVert^2} (z) \, f_{\lVert\bm{h}_r\rVert^2} \left({\lambda}/{z}\right)\,\, d\,\mathrm{ln}(z)\\
&= \dfrac{2}{ b_r b_t \, \Gamma(c_r) \Gamma(c_t)} \, K_{c_r - c_t} \Biggl(2\sqrt{\dfrac{\lambda}{b_r b_t}}\,\Biggr)
\left(\dfrac{\lambda}{b_r b_t}\right)^{\tfrac{c_t + c_r}{2} - 1}\label{eq:cap:csit0}
\end{align}
for $\lambda > 0$, where $c_r := rm_r$, $c_t := tm_t$, $b_r := {\Omega_r}/{m_r}$, $b_t := {\Omega_t}/{m_t}$ and
$K_{\nu}(\cdot)$ is the $\nu$-th order Bessel function of the second-kind~\cite{tableofintegrals}.

Since the receiver noise is normalized (see~\eqref{eq:sys_model}) and $\mathrm{rank}(\textbf{H}) = 1$, we define the average transmit signal-to-noise ratio as $\mathrm{SNR}:= P_{\mathrm{avg}}$. In the next section, we focus on the capacity of this channel in the asymptotically low-SNR regime, i.e., $\mathrm{SNR} \to 0$. For the purpose of simplification in the low-SNR asymptotic analysis, we will often compare two functions, say $f(x)$ and $g(x)$, where $x$ approaches zero and conclude $f(x) \approx g(x)$ if they
are close. Precisely, we will use the following definition:
\begin{definition}\label{eq:def:approx}
$f(x) \approx g(x)\,\,$ if and only if $\,\,\,\,\lim\limits_{x \to 0} \,\, \dfrac{f(x)}{g(x)} = 1$.
\end{definition}

\vspace{0.1em}
\section{Low-SNR Capacity of the Keyhole MIMO Channel\\ with Perfect Channel knowledge at both Transmitter and Receiver}\label{sec:three}
\vspace{0.25em}
\subsection{Asymptotic (Low-SNR) Capacity Results}
\vspace{0.25em}
Continuing from the capacity expression in~\eqref{eq:cap:csit_do}, we recall that the optimal power distribution over a scalar fading channel $\boldsymbol{\lambda}$ is the closed-form \emph{`waterfilling formula'} expressed as $P(\lambda) = \left(\tfrac{1}{\lambda_0} - \tfrac{1}{\lambda}\right)^+$~\cite{goldsmith}. Here, $z^{+}$ denotes $\mathrm{max}\{0,z\}$. The parameter $\lambda_0$ is chosen to meet the power constraint~\eqref{eq:power_const_a}. Substituting $P(\lambda)$ in~\eqref{eq:cap:csit_do}, we get
\begin{align}
C &= \int_{\lambda_0}^{\infty} \log \left({\lambda}/{\lambda_0}\right) f_{\boldsymbol{\lambda}}(\lambda)  d \lambda \label{eq:cap_der}\\[0.5em]
&= \int_{\mu_0}^{\infty} \log ({{\lambda}}/{\mu_0}) f_{\boldsymbol{\mu}}(\lambda)  d \lambda \label{eq:cap_der_do}
\end{align}
where, for convenience, we have defined a scaled random variable $\boldsymbol{\mu} := \dfrac{\boldsymbol{\lambda}}{b_r b_t}$ with the distribution as follows:
\begin{align}\label{eq:taildisb0}
f_{\boldsymbol{\mu}} (\lambda) &= \dfrac{2}{ \Gamma(c_r) \, \Gamma(c_t)}
\,\,\lambda^{\tfrac{c_t + c_r}{2} - 1} \,\cdot\, K_{c_r - c_t} \left(2\sqrt{\lambda}\right),\,\,\lambda > 0.
\end{align}
Accordingly, the power constraint~\eqref{eq:power_const_a} in terms of $\boldsymbol{\mu}$ becomes
\begin{align}\label{eq:cutoff1}
\mathrm{SNR} \left(b_t  b_r\right) =  \int_{\mu_0}^{\infty}\left( \dfrac{1}{\mu_0} -\dfrac{1}{\lambda}\right) f_{\boldsymbol\mu}(\lambda) d \lambda.
\end{align}
Note that $\mu_0:= {\lambda}_0/\left(b_t b_r \right)$. It is easy to verify from~\eqref{eq:cutoff1} that as $\textrm{SNR} \to 0$, the threshold $\mu_0 \to \infty$. The low-SNR asymptotic capacity formula is stated next.
\begin{theorem}\label{eq:theorem}
For the keyhole MIMO channel with perfect instantaneous channel knowledge at the transmitter and receiver as described by~\eqref{eq:sys_model} and subjected to IID Nakagami-$m$ fadings with parameters $(m_t,\Omega_t)$ and $(m_r,\Omega_r)$ for the transmitter-to-keyhole and keyhole-to-receiver side respectively, the low-SNR capacity is given by
\begin{align}\label{eq:thm_upper_part}
&C \approx \begin{cases}
\dfrac{n^2 \mathrm{SNR} }{4}\left(\dfrac{\Omega_t \Omega_r}{ m_t  m_r}\right) \, W_{0}^{2} \left(\mathrm{SNR}^{\tfrac{-1}{n}}\right),\,\,\,\,\,\,\,\,\mathrm{if} \,\,\,\,\,n > 0,\\[1.2em]
\dfrac{\mathrm{SNR}}{4}\left(\dfrac{\Omega_t \Omega_r}{ m_t  m_r}\right) \log^2 \left(\dfrac{1}{\mathrm{SNR}}\right),\,\,\,\phantom{xx}\,\,\,\,\,\,\,\,\mathrm{if} \,\,\,\,\, n = 0,\\[1.3em]
\dfrac{n^2 \mathrm{SNR}}{4} \left(\dfrac{\Omega_t \Omega_r}{ m_t  m_r}\right) \, W_{-1}^{2} \left(-\mathrm{SNR}^{\tfrac{-1}{n}}\right),\,\,\mathrm{if} \,\,\,\,\, n < 0.
\end{cases}
\end{align}
\vspace{-0.151em}
\begin{align}\label{eq:thm_lower_part}
\,\,\,\,\,\approx \, \left(\dfrac{\Omega_t \Omega_r}{ m_t  m_r}\right) \dfrac{\mathrm{SNR}}{4}  \log^2 \left(\dfrac{1}{\mathrm{SNR}}\right),\phantom{xxxxxxxxxxxxxxxxx}
\end{align}
where $n = \frac{9}{2} - (t m_t + r m_r)$, and $W_0(\cdot)$ and $W_{-1}(\cdot)$
are the principal and the lower branches of the Lambert $W$ function, respectively.
\end{theorem}
\begin{IEEEproof}
Recall that as $\mathrm{SNR} \to 0$, $\mu_0 \to \infty$ (or equivalently $\lambda_0 \to \infty$). Thus, we apply the series expansion
for the modified Bessel function of second kind at infinity given below~\cite{tableofintegrals}:
\begin{align}\label{eq:bessel_approx}
K_v (z)  \, \approx \, \sqrt{\dfrac{\pi}{2z}} \,e^{-z} + o\left(\dfrac{1}{z}\right), \,\,\,\,\,z \to \infty.
\end{align}
in the distribution function given in~\eqref{eq:taildisb0}, which is then substituted in~\eqref{eq:cap_der_do} to give
\begin{align}
C \,&\approx\, \dfrac{\sqrt{\pi}}{\Gamma(c_t)\Gamma(c_r)} \int_{\mu_0}^{\infty} \log \left(\dfrac{{\lambda}}{\mu_0}\right) \,\lambda^{\frac{c_t + c_r}{2} - \frac{5}{4}} \, e^{-2\sqrt{\lambda}}\,\, \mathrm{d} \lambda.\label{eq:cap_der_do222}
\end{align}
To simplify~\eqref{eq:cap_der_do222}, we apply the identity given below~\cite{tableofintegrals}:
\begin{align}
\int_{a}^{\infty} \log \left(\frac{z}{a}\right) z^{b}  e^{-2\sqrt{z}} \mathrm{d} \lambda
 \,=\, 4^{-b}{G_{2,3}^{3,0}\left({2\sqrt{a}}
\middle\vert\hspace{-0.5em}
\begin{array}{c}
1,1\\
0,0,2(1+b)\\
\end{array}\hspace{-0.5em}\right)}\label{eq:cap_der_do3}
\end{align}
where $G_{p, \,q}^{m,\,n} (\cdot)$ is the  Meijer's $G$-function~\cite{tableofintegrals}. Then, taking only the first largest term in the power series expansion of the Meijer's $G$-function at input infinity as given below~\cite{tableofintegrals}:
\begin{align}
G_{2,3}^{3,0}\left({2\sqrt{a}}
\middle\vert\hspace{-0.5em}
\begin{array}{c}
1,1\\
0,0,2(1+b)\\
\end{array}\hspace{-0.5em}\right) \, \approx \, a^{b-1} \, e^{-2\sqrt{a}} \, 4^{b}
\left(  a +  \dfrac{(2 + 8b)\sqrt{a}}{4}   + \dfrac{(12 b^2 - 1)}{4} +  \dfrac{1}{4^b}\, o\left(\dfrac{1}{a}\right)^{\frac{3}{2}}\right),
\end{align}
\vspace{-1em}
\begin{flalign}
\text{we obtain}&\phantom{xxxxxxxxxxxxxxxxxxxxxxx}C \, \approx \, \dfrac{\sqrt{\pi}}{\Gamma(c_t)\Gamma(c_r)} {\mu_0}^{\frac{c_t + c_r}{2} - \frac{5}{4}} \, e^{-2\sqrt{\mu_0}}.&\label{eq:cap_der_do4}
\end{flalign}

To express the capacity in~\eqref{eq:cap_der_do4} explicitly in terms of $\mathrm{SNR}$, we will extract the $\mu_0$-$\mathrm{SNR}$ dependence in the low-SNR regime from the average power constraint~\eqref{eq:cutoff1}. To begin with, we employ the distribution~\eqref{eq:taildisb0}, with low-SNR approximation~\eqref{eq:bessel_approx} applied, in the power constraint~\eqref{eq:cutoff1} to obtain
\begin{align}\label{eq:cutoff121}
\mathrm{SNR} \left(b_t  b_r\right) \,&\approx \, \dfrac{\sqrt{\pi}}{\Gamma(c_t)\Gamma(c_r)} \left[\, \dfrac{1}{\mu_0} I_1 (\mu_0) - I_2 (\mu_0) \, \right]
\end{align}
\vspace{-1em}
\begin{flalign}\label{eq:I_definitions}
\text{where}\,\,\,\,\,\,\,\phantom{xxxxxxxxxxxxxxxxxxxx}&\begin{cases}\,
I_1 (\mu_0) \,=\, \dfrac{1}{2^{(c_t + c_r)-\frac{3}{2}}}\Gamma\bigl(c_t + c_r - \frac{1}{2}, 2\sqrt{\mu_0}\bigr),\\[0.95em]
\,I_2 (\mu_0) \,=\, \dfrac{1}{2^{(c_t + c_r)-\frac{7}{2}}}\Gamma\bigl(c_t + c_r - \frac{5}{2}, 2\sqrt{\mu_0}\bigr),
\end{cases}&
\end{flalign}
where, in turn, $\Gamma(\cdot,\cdot)$ is the upper incomplete Gamma function. Using the first two
largest terms in the series expansion of $\Gamma(a,x)$ function with input $x$ approaching infinity
given below:
\begin{align}\label{eq:gamma_approx}
\Gamma(a,x) \, \approx \, e^{-x} x^a \left(\dfrac{1}{x} + \dfrac{a-1}{x^2} + o\left(\dfrac{1}{x}\right)^3\right),
\end{align}
the average power constraint in~\eqref{eq:cutoff121} gets simplified as
\begin{align}\label{eq:cutoff131}
\mathrm{SNR} \left({b_t  b_r}\right) \,&\approx\, \dfrac{\sqrt{\pi}}{\Gamma(c_t)\Gamma(c_r)}  {\mu_0}^{\frac{c_t + c_r}{2} - \frac{9}{4}} \, e^{-2\sqrt{\mu_0}}.
\end{align}
Comparison of~\eqref{eq:cutoff131} and~\eqref{eq:cap_der_do4} implies $C \, \approx \, \mu_0 (b_t  b_r)\, \mathrm{SNR}$ or simply $C \, \approx \, \lambda_0 \mathrm{SNR}$.
Notice that~\eqref{eq:cutoff131} can be expressed in the form of $y \,=\,  x e^x$
which, in turn, can be solved using the principal and the lower branches of the Lambert $W$ function
depending on the value of $\,n = \tfrac{9}{2} - (c_t + c_r)$. With~\eqref{eq:cutoff131} rewritten in the
$y =  x e^x$ form as
 \begin{align}\label{eq:cutoff1313}
\dfrac{2}{n} ( \tau \mathrm{SNR})^{-\frac{1}{n}} \,=\, \dfrac{2\sqrt{\mu_0}}{n} \, e^{\frac{2\sqrt{\mu_0}}{n} }
\end{align}
where $\tau \,=\, \dfrac{\Gamma(c_t)\Gamma(c_r) \left({b_t  b_r}\right)}{\sqrt{\pi}}$, we now solve for $\mu_0$ as follows:
\begin{itemize}
\item If $n \,=\, 0$, then~\eqref{eq:cutoff131} simplifies to $\tau \mathrm{SNR} \approx
e^{-2\sqrt{\mu_0}}$ which is solved as
\begin{align}
\mu_0 \,&\approx\,  \dfrac{1}{4} \log^2 \left(\dfrac{1}{\tau \mathrm{SNR}}\right)\label{eq:n_zero1}\\[0.45em]
      \,&\approx\,  \dfrac{1}{4} \log^2 \left(\dfrac{1}{\mathrm{SNR}}\right),\label{eq:n_zero2}
\end{align}
where the $\tau$ parameter (see~\eqref{eq:n_zero1}) is neglected in~\eqref{eq:n_zero2} by applying the Definition~\ref{eq:def:approx} using the log-function limit property that $\lim\limits_{z \to \infty} \dfrac{\log (\theta z)}{\log (z)} \,=\, 1$ for any $\theta \,>\, 0$.
\item If $n \,>\, 0$, then~\eqref{eq:cutoff1313} is solved using the principal branch of the
Lambert $W$ function $W_0 (\cdot)$ since $\dfrac{2\sqrt{\mu_0}}{n} \,>\, 0$, to give
\begin{align}\label{eq:lamda_n_more}
\mu_0 \,\approx\, \left[ \dfrac{n}{2} \, W_{0} \left(\dfrac{2}{n} ( \tau \mathrm{SNR})^{-\frac{1}{n}} \right) \right]^2 .
\end{align}
Applying the Definition~\ref{eq:def:approx} using the property that $\lim\limits_{z \to \infty} \dfrac{W_0 (\theta z) }{W_0 (z)} \,=\, 1$ for any $\theta \,>\, 0$, we have
\begin{align}\label{eq:n_below0}
\mu_0 \, \approx \, \dfrac{n^2}{4} \, W_{0}^{2} \left(\left(\dfrac{1}{\mathrm{SNR}}\right)^{\frac{1}{n}}\right).
\end{align}
\item If $n \, < \, 0$, then~\eqref{eq:cutoff1313} is solved using the lower branch of the Lambert $W$ function $W_{-1} (\cdot)$ since $\dfrac{2\sqrt{\mu_0}}{n} \, < \, 0$, to give
\begin{align}\label{eq:n_below}
\mu_0 \, \approx \, \left[ \dfrac{n}{2} \, W_{-1} \left(\dfrac{2}{n} ( \tau \mathrm{SNR})^{-\frac{1}{n}} \right) \right]^2.
\end{align}
Now, we apply the Definition~\ref{eq:def:approx} using the property that $\lim\limits_{z \to 0+} \dfrac{W_{-1} (\theta z) }{W_{-1} (-z)} \,=\, 1$ for any $\theta \,<\, 0$ to have
\begin{align}\label{eq:n_below1}
\mu_0 \, \approx \, \dfrac{n^2}{4} \, W_{-1}^{2} \left(-\left(\dfrac{1}{\mathrm{SNR}}\right)^{\frac{1}{n}}\right).
\end{align}
\end{itemize}
Finally, rewriting $C  \, \approx \, \mu_0 \left({b_t  b_r}  \right) \mathrm{SNR}$ with $\mu_0$ (expressed in terms of $\mathrm{SNR}$) in~\eqref{eq:n_zero2},~\eqref{eq:n_below0},~\eqref{eq:n_below1}, completes the
proof of~\eqref{eq:thm_upper_part} in Theorem~\ref{eq:theorem}.

Alternatively, we can express the asymptotic ergodic capacity in a simple $\log(\cdot)$ function form as given in~\eqref{eq:thm_lower_part} by taking the logarithm on both sides of~\eqref{eq:cutoff131} and neglecting smaller terms (in magnitude) to obtain
\begin{align}\label{eq:n_below1223}
\log (\mathrm{SNR}) \, \approx \, -2 \sqrt{\mu_0}.
\end{align}
Substituting $\mu_0$ from~\eqref{eq:n_below1223} into $C \, \approx \, \mu_0 \left({b_t  b_r}  \right) \mathrm{SNR}$ completes the proof of~\eqref{eq:thm_lower_part}.
\end{IEEEproof}

\begin{figure*}[!b]
\centering
\begin{subfigure}{.5\linewidth}
\includegraphics[scale = 0.75]{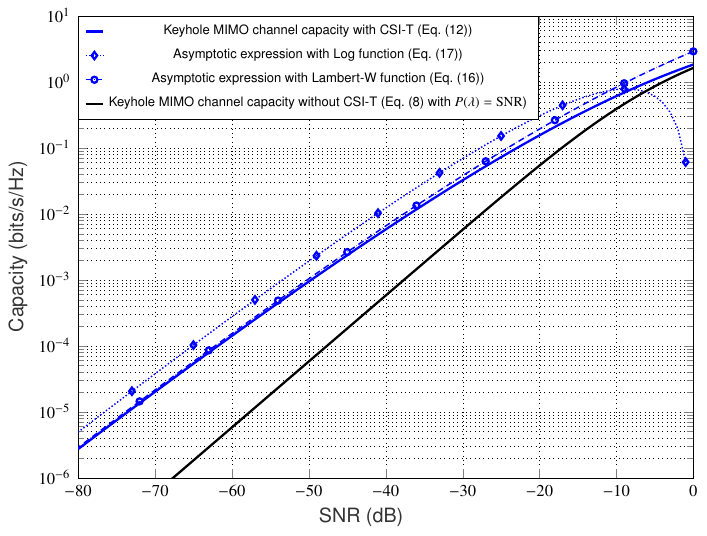}
\caption{}
\label{fig:one}
\end{subfigure}%
\begin{subfigure}{.5\linewidth}
\centering
\includegraphics[scale = 0.75]{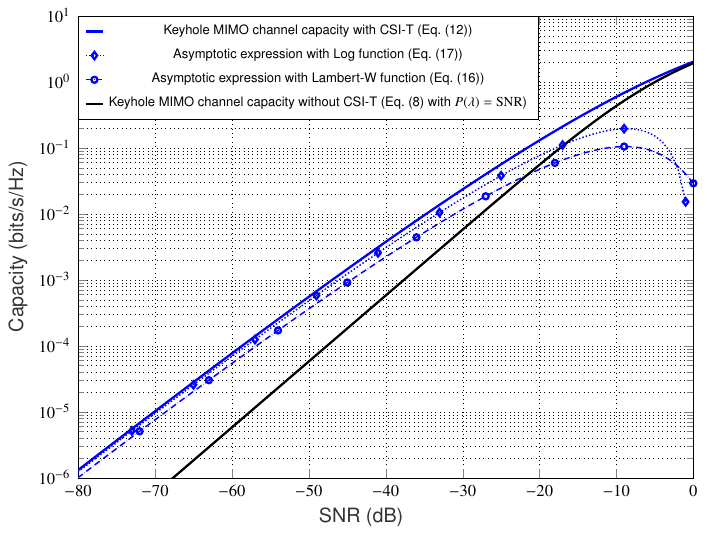}
\caption{}
\label{fig:two}
\end{subfigure}
\vspace{2em}
\newline
\begin{subfigure}{.495\linewidth}
  \centering
  \includegraphics[scale = 0.75]{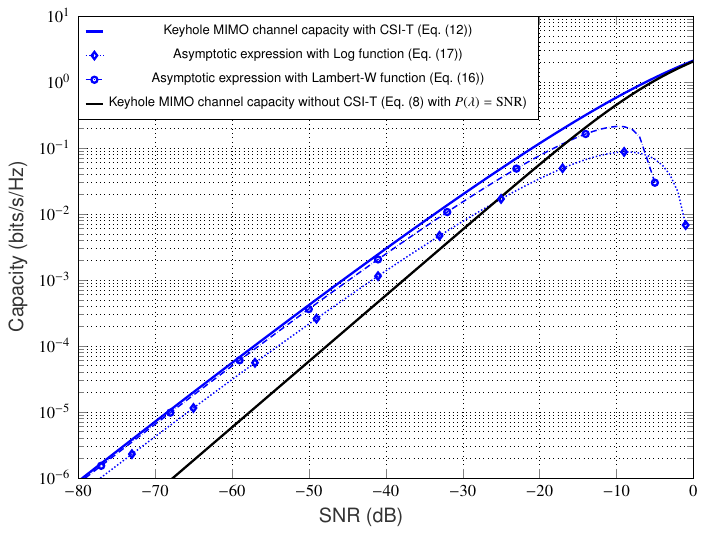}
\caption{}
\label{fig:three}
\end{subfigure}
\begin{subfigure}{.495\linewidth}
  \centering
  \includegraphics[scale = 0.75]{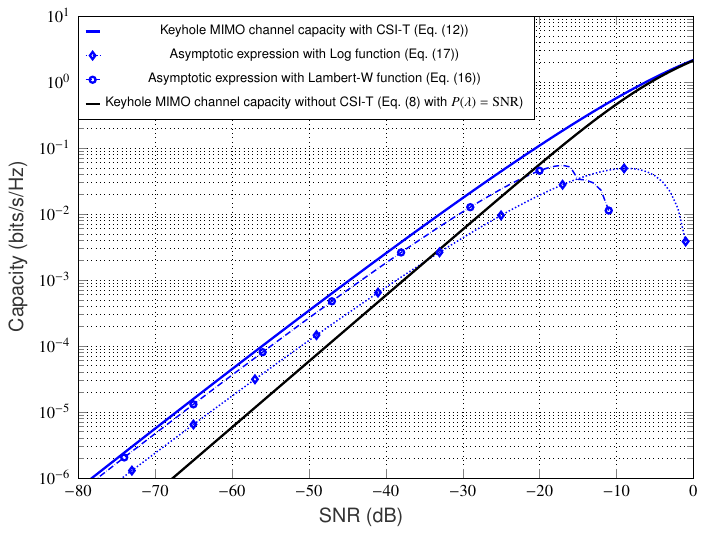}
\caption{}
\label{fig:four}
\end{subfigure}
\vspace{2em}
\caption{Low-SNR capacity of $2 \times 2$ keyhole MIMO channel in IID normalized Nakagami-$m$ fading conditions:
(a) $m_r = m_t = \tfrac{1}{2}$, $\Omega_r = \Omega_t = 1$ $\Rightarrow$ $n = \tfrac{5}{2}$, (b) $m_r = m_t = 1$, $\Omega_r = \Omega_t = 1$ $\Rightarrow$
$n = \tfrac{1}{2}$, (c) $m_r = m_t = \tfrac{3}{2}$, $\Omega_r = \Omega_t = 1$ $\Rightarrow$ $n = -\tfrac{3}{2}$, and (d) $m_r = m_t = 2$, $\Omega_r = \Omega_t = 1$ $\Rightarrow$ $n = -\tfrac{7}{2}$.}
\end{figure*}
To illustrate the accuracy of the asymptotic low-SNR capacity formulae proposed in Theorem~\ref{eq:theorem}, we now present some numerical results. The exact (i.e., non-asymptotic) capacity curves with and without channel knowledge at the transmitter (for reference/comparison purpose) are computed by standard numerical integration methods; the  threshold $\lambda_0$ is  also computed numerically from the average power constraint. For simplicity, we have normalized all the fading gains to unity, i.e., $\Omega_r = \Omega_t = 1$. The channel settings in Figures~\ref{fig:one} and~\ref{fig:two} corresponds to $n > 0$ case, and in Figures~\ref{fig:three} and~\ref{fig:four} corresponds to $n < 0$ case. From these Figures, it is clear that the curves of the asymptotic capacity expressions in Theorem~\ref{eq:theorem} follow the same shape as of the exact capacity curves in the displayed SNR range. In both cases, we have verified that by further reducing the $\mathrm{SNR}$ considerably, the gap to the exact capacity reduces significantly. Notice from the Figures~\ref{fig:one} and~\ref{fig:two} that at sufficiently low SNR values, the Log function based characterization of the asymptotic capacity in~\eqref{eq:thm_lower_part} is always an upper bound on the Lambert $W$ function based characterization in~\eqref{eq:thm_upper_part} for $n > 0$; likewise, from the Figures~\ref{fig:three} and~\ref{fig:four}, we note that~\eqref{eq:thm_lower_part} is always a lower bound on~\eqref{eq:thm_upper_part} at low SNRs for $n < 0$ (see Appendix~\ref{sec:app:A} for the detailed proofs). We have also observed that both of these asymptotic capacity characterizations are better for $n$ values close to zero.

\subsection{On-Off Transmission Policy Is Asymptotically Optimal}\label{sec:3B}
The ergodic rate performance of an on-off power scheme has recently been explored for a few important fading channels and is found to be asymptotically optimal in the low-SNR regime~\cite{tall}-\cite{benkhelifa2}. The attractive feature of this transmission scheme is that it requires feedback of only one-bit channel information (i.e., whether current channel state is in good or bad condition) at the transmitter for power adaptation. We will explore the ergodic rate performance of the on-off transmission scheme over the keyhole MIMO channel in the low-SNR conditions and show that the on-off transmission rates follows the keyhole MIMO channel capacity closely at low-SNRs. The mathematical details are relegated to Appendix~\ref{sec:app:B}.

\section{Impact of Keyhole Degeneracy\\ on the MIMO Channel capacity in the Low-SNR Regime}\label{sec:four}
\vspace{0.1em}

In this section, we will compare and analyze the low-SNR capacity behaviour of the MIMO fading channels with and without degenerate keyhole condition. When the SNR is high, it is well-known that the capacity of MIMO fading channel degrades severely in the presence of keyhole effect due to reduction of the spatial degrees of freedom to unity~\cite{gesbert}-\cite{shinlee}. On the other hand, in the low-SNR regime, the rank of the pure MIMO fading channel has very limited or little impact on the ergodic capacity~\cite{tse}; this holds for keyhole MIMO channel by default (and independent of SNR) since channel rank is unity. Hence, with channel rank no longer a dominanting factor, it will be interesting to analyze the low-SNR capacity behaviour of the MIMO fading channel with and without keyhole effect.

To simplify the exposition and derive some useful insights for performance comparison, we will first ignore large-scale attenuation (or path loss) effects and consider only small scale fading effects on the capacity of the MIMO fading channel with and without keyhole degeneracy. However, for an accurate and realistic comparison, it is \emph{`necessary'} to incorporate suitable large-scale fading laws between the MIMO transmitter and receiver sides with and without keyhole condition. The reason is that, similar to the fact that  small-scale fading characteristics vary significantly when keyhole condition is introduced, the large-scale attenuation law also gets altered significantly under keyhole effect. For analytical convenience, we will henceforth consider the classical IID Rayleigh fading MIMO channel with and without keyhole effect and under perfect channel knowledge assumption at the transmitter and receiver sides. Furthermore, we will assume normalized mean Rayleigh fading gains. Thus, $m_r = m_t = 1$ and $\Omega_r = \Omega_t = 1$.
\subsection{Low-SNR MIMO Channel Capacity under Keyhole Effect Without Large-scale Attenuation (Path Loss) Considerations}\label{sec:4A}
Under no path loss considerations and only normalized small-scale Rayleigh fadings present, the low-SNR capacity of the MIMO fading channel (no keyhole) follows asymptotically as~\cite[Eq.~(6)]{tall}
\begin{align}\label{eq:mimolowSNR}
\mathrm{SNR} \log{\dfrac{1}{\mathrm{SNR}}}.
\end{align}
When subjected to keyhole condition, Theorem~\ref{eq:theorem} suggests that the low-SNR capacity of this MIMO fading channel scales asymptotically as
\pagebreak
\begin{align}\label{eq:keyholemimolowSNR}
\dfrac{\mathrm{SNR}}{4}  \log^2 \left(\dfrac{1}  {\mathrm{SNR}}\right).
\end{align}
\begin{figure*}[!b]
\centering
\includegraphics[scale = 1.37   ]{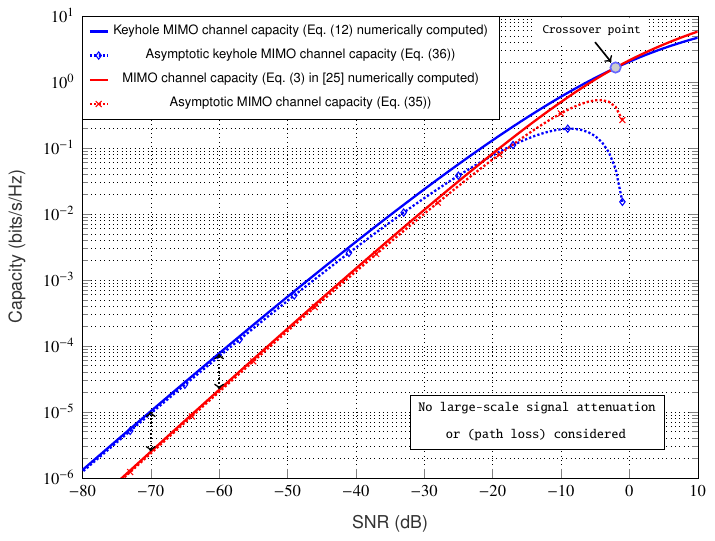}
\vspace{1em}
\caption{Capacity of $2 \,\times\, 2$ MIMO IID Rayleigh channel with and without keyhole effect. Note that the \emph{crossover point} lies approximately at $\mathrm{SNR} = -2$ $\mathrm{dB}$.}
\label{fig:sixA}
\end{figure*}
It is now straightforward to conclude that~\eqref{eq:keyholemimolowSNR} exceeds~\eqref{eq:mimolowSNR} in the low-SNR regime, see Fig.~\ref{fig:sixA} for numerical illustration. Precisely,~\eqref{eq:mimolowSNR} is improved by a factor of
\begin{align}\label{eq:improve1}
\dfrac{1}{4}  \log \left(\dfrac{1}{\mathrm{SNR}}\right)
\end{align}
when subjected to keyhole condition which improves unboundedly as $\mathrm{SNR} \to 0$; this is depicted in Fig.~\ref{fig:sixA} by black vertical double arrows for the $2 \,\times\, 2$ MIMO IID Rayleigh channel in the low-SNR regime. Notice that the exact capacity of the MIMO Rayleigh fading channel under keyhole effect begins to exceed the capacity pure MIMO Rayleigh fading channel starting at $\mathrm{SNR}$ close to $-2$ $\mathrm{dB}$. We define this event as \emph{`crossover phenomenon'} and the respective $\mathrm{SNR}$ is termed as \emph{`crossover point'}. To justify this improvement, note that the `spatial degrees of freedom gain'  is no longer a dominant factor for the MIMO fading channel (with and without keyhole) in the low-SNR regime. On the other hand, a keyhole MIMO channel is double-faded and thus more severe than the MIMO fading channel without keyhole effect. With this increased fading severity, the knowledge of the frequent channel fluctuations (e.g., peaks and nulls) at the transmitter helps to improve the performance in the low-SNR regime by dynamically allocating power and transmitting \textit{opportunistically} only when the channel state is good. The low-SNR capacity formulas derived recently for a few important scalar fading channels such as Nakagami-$m$ fading~\cite{rezki}, Gamma-Gamma atmospheric turbulence fading~\cite{benkhelifa}, and Rayleigh-product fading~\cite{benkhelifa2} also clearly allude to our observation of enhanced low-SNR capacity with higher fading severity levels. It is expected that with the Nakagami-$m$ fading shape parameter $m$ increasing (i.e., fading severity level is decreasing), the capacity enhancement of the keyhole MIMO channel in the low-SNR regime narrows down (when compared to MIMO channel without keyhole) as there are lesser opportunistic communications possible due to relatively mild channel fluctuations.

\subsection{Low-SNR MIMO Channel Capacity under Keyhole Effect With Large-scale Attenuation (Path Loss) Considerations}\label{sec:4B}
The low-SNR capacity improvement of MIMO fading channel in the presence of keyhole (as indicated in Fig.~\ref{fig:sixA}) is \emph{over-simplified} due to the lack of suitable large-scale attenuation (or path loss) laws. It is important to note that path loss in  wireless propagation conditions diminishes the received signal power very rapidly.
As we will discuss later that large-scale attenuation (path-loss) laws are significantly different for propagation conditions with and without keyhole effect. More precisely, the large-scale signal attenuation (path loss) increases significantly when signal propagation is subjected to keyhole condition. Therefore, the performance improvement suggested in Fig.~\ref{fig:sixA} is \emph{overly optimistic} and \emph{unrealistic}. Nevertheless, this simplistic analysis  provides a critical insight that the improvement factor (see~\eqref{eq:improve1}) increases unboundedly as $\mathrm{SNR}$ gets smaller which, in turn, suggests that at sufficiently low-SNR, the keyhole MIMO channel capacity should surpass the pure MIMO channel capacity irrespective of the huge difference in path losses. Motivated by this insight, we will now perform a detailed low-SNR capacity analysis of keyhole MIMO channels under more realistic propagation conditions. We will incorporate suitable large-scale attenuation (path loss) laws along with the small-scale IID Rayleigh fadings in the MIMO channel with and without keyhole condition. To account for the large-scale attenuation, we employ a widely used distance-dependent path loss model where the average received signal power decays logarithmically with distance. This observation is theoretically-consistent and also supported quite well by emperical measurements for both indoor and outdoor wireless channels~\cite{rappaport}. Assuming that the transmit power is $P_T$, the average received signal power is as follows:
\begin{align}\label{eq:powernoKeyhole}
P_R = K_{mimo} \left(\dfrac{P_T}{d_{TR}^\gamma}\right)
\end{align}
which decays with the $\gamma$-th power of the distance $d_{TR}$ between the transmitter and receiver sides of the MIMO channel, and the proportionality constant $K_{mimo}$ is determined from measurements at some close-in free space reference distance. For example, with $\lambda$ as the operating wavelength, $K_{mimo} = (\lambda/4\pi)^2$ is the free space path loss at reference distance of $1 \, \textrm{m}$. The value of $\gamma$, also known as path loss exponent, is dependent on the specific signal propagation conditions; e.g., the value $\gamma$ equals $2$ for line of sight propagation (or free space propagation) and larger value for more difficult propagation conditions~\cite[Chapter~3]{rappaport}.

The channel matrix for the MIMO IID Rayleigh channel with large-scale fading incorporated and \emph{without} keyhole condition can be expressed as
\begin{align}\label{eq:MimoHpath loss}
\textbf{H}_{mimo} := \sqrt{\varrho} \,\, \textbf{H}_{Rayleigh}
\end{align}
where the constant $\varrho$ is the large-scale attenuation corresponding to
the total distance $d_{TR}$ between the multi-antenna transmitter and receiver, i.e., $\varrho := {P_R/P_T}= {K_{mimo}/d_{TR}^\gamma}\,$. The $r \,\times\, t$ Rayleigh channel matrix $\textbf{H}_{Rayleigh}$ has normalized IID Rayleigh fading (small-scale) entries, i.e., each entry in $\textbf{H}_{Rayleigh} \sim\mathcal{CN}(0,1)$. We will borrow the optimal power and capacity laws for the MIMO channel $\textbf{H}_{Rayleigh}$ from~\cite{kamal_itit} and scale these laws appropriately for $\textbf{H}_{mimo}$. The ergodic capacity integral formula for $\textbf{H}_{Rayleigh}$ is available in~\cite[Eq.~(2)]{kamal_itit} and can be appropriately altered for $\textbf{H}_{mimo}$ as follows:
\begin{align}
C_{mimo} &= m \, \mathbb{E}_{{{\boldsymbol{\zeta}}}} [\log(1 \,+\, \varrho\, {\boldsymbol{\zeta}} \, {P}^{\,*}({\boldsymbol{\zeta}}))] \label{eq:Cap:nokeyholenewone111} \\[0.65em]
&= m \int_{\zeta_0}^{\infty} \log\left(\dfrac{\varrho {\zeta}}{\zeta_0}\right)  f_{{\boldsymbol{\zeta}}}(\zeta) d \zeta \label{eq:Cap:nokeyholenewone222} \\
  &= m (1 - F_{\bar{\boldsymbol{\zeta}}}(\zeta_0)) \log \varrho  + m \int_{\zeta_0}^{\infty} \log\left(\dfrac{{\zeta}}{\zeta_0}\right)   f_{{\boldsymbol{\zeta}}}(\zeta) d \zeta \label{eq:Cap:nokeyholenewone}
\end{align}
where $m := \mathrm{min}\{r,t\}$ and the second equality follows from the substitution of the optimal waterfilling scheme ${P}^{\,*}(\zeta) = \left(\dfrac{1}{\zeta_0}-  \dfrac{1}{\varrho \zeta} \right)^+$. The \emph{eigenmode} distribution $f_{{\boldsymbol{\zeta}}} (\zeta)$ is given by
\vspace*{7pt}
\begin{align}\label{eq:eigendist}
f_{{\boldsymbol{\zeta}}} (\zeta) = \dfrac{e^{-\zeta} \zeta^{n-m} }{m} \,\, \sum_{k=0}^{m-1} \, \dfrac{k!}{(k+n-m)!} \, [L_{k}^{n-m}(\zeta)]^2\\[-13pt] \notag
\end{align}
where, in turn, $n := \mathrm{max}\{r,t\}$ and $L_k^{n-m}(\zeta)$, the associated Laguerre polynomial of order $k$, has a closed-form expression given as
\begin{align}\label{eq:associatedLaguarre:simplification}
L_{k}^{n-m}(\zeta) = \sum_{p = 0}^{k} \dfrac{(-\zeta)^{p}}{p!} {k + n - m \choose k - p} \cdot
\end{align}
The corresponding average power constraint is obtained by modifying from~\cite[Eq.~(4)]{kamal_itit} as follows:
\begin{align}\label{eq:Power:keyholenewone12}
\dfrac{\varrho \, \mathrm{SNR}}{m} =  \int_{\zeta_0}^{\infty}\left(\dfrac{1}{\zeta_0} -\dfrac{1}{\zeta}\right) f_{{\boldsymbol\zeta}}(\zeta) d \zeta.
\end{align}
As mentioned earlier, the low-SNR capacity of the MIMO IID Rayleigh channel $\textbf{H}_{Rayleigh}$ has been investigated in~\cite{tall} and is shown to follow (asymptotically) as
\begin{align}
\mathrm{SNR} \log{\dfrac{1}{\mathrm{SNR}}}.
\end{align}
Keeping~\eqref{eq:Cap:nokeyholenewone} and~\eqref{eq:Power:keyholenewone12} in mind for the channel $\textbf{H}_{mimo}$ and following similar steps as in the proof of the main asymptotic capacity result in Section~III in~\cite{tall}, it can be easily shown that the low-SNR capacity of the channel $\textbf{H}_{mimo}$ scales asymptotically as
\begin{align}\label{eq:asympnoKeyholelossy}
C_{mimo} \, \approx \, \mathrm{\varrho SNR}  \log \left(\dfrac{1} {\mathrm{ \varrho SNR}}\right).
\end{align}

For the MIMO fading channel under the keyhole condition, the distance-dependent path loss model results in the average received signal power as follows:
\begin{align}\label{eq:powerKeyhole}
P_R = K_{keyhole} \left(\dfrac{P_T}{d_{TK}^\gamma d_{KR}^\gamma}\right)
\end{align}
where $d_{TK}$ and $d_{KR}$ are the distance from the keyhole to the transmitter and receiver, respectively. The proportionality constant $K_{keyhole}: = \lambda^2 A/ (4 \pi)^3$ where $A$ is the area of the keyhole. The equation~\eqref{eq:powerKeyhole} with  $K_{keyhole}$ defined earlier is well known in the literature as ``radar equation'' and, roughly speaking, it is valid when A is `small' (i.e., $A/\lambda<< \mathrm{min}(d_{TK},d_{KR})$). It is important to note that the keyhole size plays a critical role in the amount of scattering possible; a small keyhole on the order of the wavelength provides good scattering (coverage) in all directions with rank reduction while a large keyhole does not result in a rank reduction as anticipated in a keyhole effect~\cite{almers0}-\cite{almers}. It is clear from~\eqref{eq:powernoKeyhole} and~\eqref{eq:powerKeyhole} that, similar to the fact that  small-scale fading characteristics vary significantly when keyhole condition is introduced, the large-scale attenuation law is also modified significantly with wireless signal propagation subjected to keyhole effect. To give a simple example, assuming that $d_{TK} = d_{KR} = d_{TR}/2$, i.e., the keyhole is located in the middle of the channel between the transmitter and receiver sides, we obtain $P_R \propto P_T/ d_{TR}^{2\gamma}$ and thus, the path loss exponent is twice larger (i.e., more severe) than that without keyhole condition. This explains why it is imperative to incorporate large-scale attenuation laws (path losses) along with the small-scale fadings for an accurate comparison study of the MIMO fading channel capacity with and without keyhole effect. Now, we will incorporate the modified large-scale attenuation law (fixed path loss) into the optimal waterfilling power and capacity laws for the MIMO fading channel with keyhole degeneracy.

Applying the path loss scaling law in~\eqref{eq:powerKeyhole} with transmitter-to-keyhole distance as $d_{TK}$ and keyhole-to-receiver distance as $d_{KR}$, the keyhole MIMO channel in~\eqref{eq:keyholeH} is modified as
\begin{align}\label{eq:keyholeHpath loss}
\textbf{H}_{keyhole} := \sqrt{\xi} \,\, \textbf{h}_r \textbf{h}_t^{T}
\end{align}
where the constant $\xi$ is the large-scale attenuation corresponding to
the distances $d_{TK}$ and $d_{KR}$ respectively as $\xi := P_R/P_T = K_{keyhole}/\left(d_{TK}^\gamma d_{KR}^\gamma \right)$. Note that $\textbf{h}_r$ and $\textbf{h}_t$ are small-scale fadings similar to as defined in the system model in Section~\ref{sec:intro}. Accordingly, the ergodic capacity in~\eqref{eq:cap:csit_do} becomes
\begin{align}\label{eq:Cap:keyholenewone0}
C_{keyhole} &= \mathbb{E}_{\boldsymbol{\lambda}} [\log(1 + \xi \boldsymbol{\lambda} P(\boldsymbol{\lambda}))]
\end{align}
with the optimal waterfilling power adaptation $P(\lambda) = \left(\dfrac{1}{\lambda_0}-  \dfrac{1}{\xi \lambda} \right)^+$. Substituting this waterfilling power scheme in~\eqref{eq:Cap:keyholenewone0} gives
\begin{align}\label{eq:Cap:keyholenewone}
C_{keyhole} &= \int_{\lambda_0}^{\infty} \log(\xi {\lambda}/{\lambda_0})  f_{\boldsymbol{\lambda}}(\lambda) d \lambda \notag \\
  &= (1 - F_{\boldsymbol{\lambda}}(\lambda_0)) \, \log (\xi)  \,\,+ \int_{\lambda_0}^{\infty} \log\left({\lambda}/{\lambda_0}\right)  f_{\boldsymbol{\lambda}}(\lambda) d \lambda.
\end{align}
Correspondingly, the average power constraint becomes
\begin{align}\label{eq:Power:keyholenewone}
\xi \, \mathrm{SNR} =  \int_{\lambda_0}^{\infty}\left(\dfrac{1}{\lambda_0} -\dfrac{1}{\lambda}\right) f_{\boldsymbol\lambda}(\lambda) d \lambda
\end{align}
with $\mathrm{SNR}$, $\boldsymbol\lambda$ and its PDF $f_{\boldsymbol\lambda}(\cdot)$
are exactly the same as defined in Section~\ref{sec:intro}. Now keeping~\eqref{eq:Cap:keyholenewone} and~\eqref{eq:Power:keyholenewone} in mind and following steps similar to as in the proof of Theorem~\ref{eq:theorem}, it is a straightforward exercise to show that the low-SNR capacity of this keyhole MIMO Rayleigh channel $\textbf{H}_{keyhole}$ scales asymptotically as
\begin{align}\label{eq:asympKeyholelossy}
C_{keyhole} \approx \dfrac{\xi \mathrm{SNR}}{4}  \log^2 \left(\dfrac{1} {\xi \mathrm{SNR}}\right).
\end{align}


With~\eqref{eq:asympnoKeyholelossy} as the low-SNR asymptotic capacity for the pure MIMO fading channel $\textbf{H}_{mimo}$ and~\eqref{eq:asympKeyholelossy} as the low-SNR asymptotic capacity for the keyhole MIMO fading channel $\textbf{H}_{keyhole}$ under IID Rayleigh fadings condition, we now state the main result in the following theorem.
\begin{theorem}\label{eq:theorem2}
In the asymptotically low-SNR regime, the capacity of the keyhole MIMO Rayleigh channel $\textbf{H}_{keyhole}$ (described by~\eqref{eq:keyholeHpath loss}) exceeds the capacity of the pure MIMO Rayleigh channel $\textbf{H}_{mimo}$ (described by~\eqref{eq:MimoHpath loss}), i.e.,
\begin{align*}
C_{keyhole}\,\,>\,\,  C_{mimo}\,\,\,\,\textrm{as}\,\,\,\,\mathrm{SNR}\,\to\,0.
\end{align*}
\end{theorem}
\begin{IEEEproof}
Note that $\varrho < 1$, $\xi < 1$ and $\xi << \varrho$. With the low-SNR asymptotic capacity expressions in~\eqref{eq:asympnoKeyholelossy} and~\eqref{eq:asympKeyholelossy} for the MIMO Rayleigh channels $\textbf{H}_{mimo}$  and $\textbf{H}_{keyhole}$ respectively, we have
\begin{align}
\dfrac{C_{keyhole}}{C_{mimo}} \,\, &\approx \,\, \dfrac{\tfrac{\xi\,\mathrm{SNR}}{4}\log^2 \left(\tfrac{1}{{\xi\, \mathrm{SNR}}}\right)}{\varrho\,\mathrm{SNR}\log\left(\tfrac{1}{{\varrho\, \mathrm{SNR}}}\right)}\nonumber \\[0.35em]
&= \,\, \dfrac{\xi\,}{4\varrho\,} \tfrac{\log^2 \left(\tfrac{1}{{\xi\,\mathrm{SNR}}}\right)}{\log\left(\tfrac{1}{{\varrho\, \mathrm{SNR}}}\right)}\nonumber\\
&> \,\, \dfrac{\xi\,}{4\varrho\,} \tfrac{\log^2 \left(\tfrac{1}{{\varrho\, \mathrm{SNR}}}\right)}{\log\left(\tfrac{1}{{\varrho\, \mathrm{SNR}}}\right)}\nonumber\\
&= \,\, \dfrac{\xi\,}{4\varrho\,} \log \left(\dfrac{1}{{\varrho\, \mathrm{SNR}}}\right).\label{eq:lowerbound:ab}
\end{align}
Notice that the lowerbound on the capacity improvement factor in~\eqref{eq:lowerbound:ab} increases unboundedly with decreasing $\mathrm{SNR}$ for fixed $\varrho$ and $\xi$. Thus, it is guaranteed that the $C_{keyhole}$ exceeds $C_{mimo}$ at sufficiently low-SNRs.
\end{IEEEproof}

\subsection{Numerical Results and Discussion}\label{sec:4C}
\begin{figure*}[!t]
\centering
\begin{subfigure}{.5\linewidth}
\includegraphics[scale = 0.73]{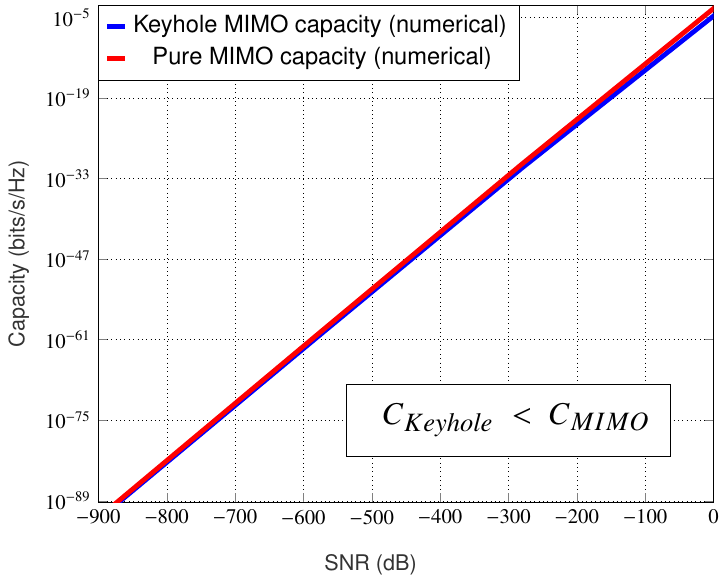}
\vspace{-1.3em}
\caption{}
\label{fig:sevenA}
\end{subfigure}%
\begin{subfigure}{.5\linewidth}
  \centering
  \includegraphics[scale = 0.73]{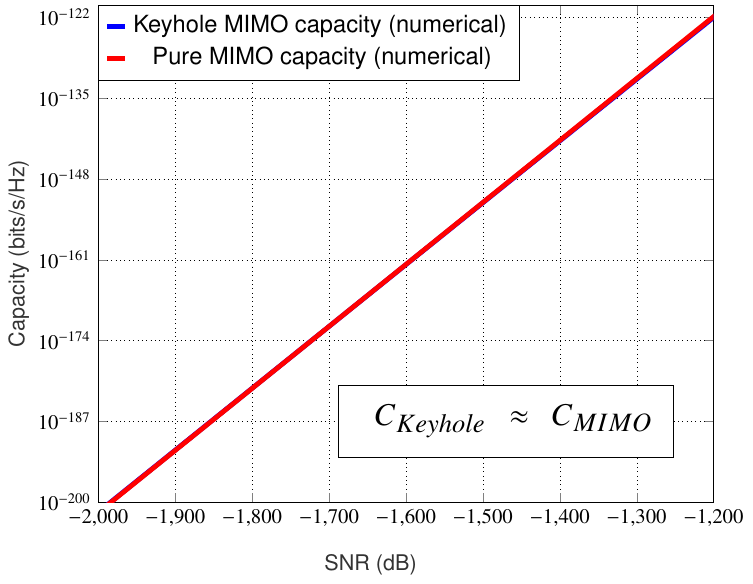}
\vspace{-1.3em}
\caption{}
\label{fig:sevenB}
\end{subfigure}
\newline
\begin{subfigure}{1\linewidth}
\vspace{1.0em}
\centering
\includegraphics[scale = 1.45]{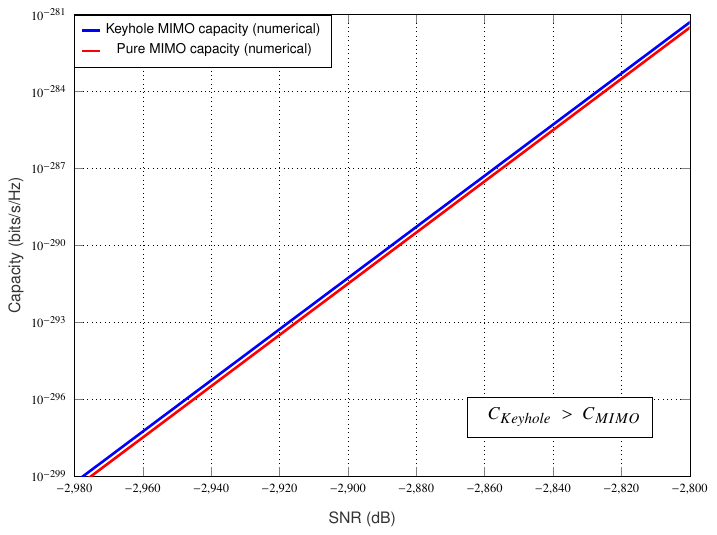}
\vspace{-1.42em}\phantom{xxxxxxxc}\caption{}
\label{fig:sevenC}
\end{subfigure}%
\vspace{1.em}
\caption{$2 \times 2$ MIMO channel with and without keyhole in IID Rayleigh fadings  with path loss models~\eqref{eq:powernoKeyhole} and~\eqref{eq:powerKeyhole} included respectively. Exact capacity variations as SNR gets smaller (a) $C_{keyhole} < C_{mimo}$ in the displayed SNR range $[0,-900]$ dB,  (b) $C_{keyhole} \approx  C_{mimo}$ in the $[-1200,-2000]$ dB SNR range, and finally (c) $C_{keyhole} > C_{mimo}$ clearly visible in the $[-2800,-2980]$ dB SNR range.}
\end{figure*}

For numerical demonstration, we assume the path loss related parameters as follows: wavelength of operation $\lambda = 3/50$ $\textrm{m}$ (equivalently RF frequency of $5$ GHz), $K_{mimo} = 10^{-4.642}$ (or $-46.42$ $\textrm{dB}$) and $\gamma = 2$. The choice $K_{mimo} = 10^{-4.642}$ or $\,46.42$ $\textrm{dB}$ attenuation can be shown to be accurate at a reference distance of $1$ meter with frequency of operation $5$ GHz and unity gain antennas. Also, the selection of the path loss exponent $\gamma = 2$ is reasonable for microcellular radio operation~\cite{rappaport}. The keyhole area is taken as $A = \lambda^2$ which can be considered a relatively larger hole (see~\cite{almers0}-\cite{almers} for keyhole's size and measurements related experimental work). For these settings, the keyhole channel's path loss constant $K_{keyhole} = -81.85$ $\textrm{dB}$. Also, we take distances $d_{TK} = 10 \lambda$, $d_{KR} = 4 \lambda$ and $d_{TK} + d_{KR} \approx d_{TR}$ (thus, $d_{TR} = 14 \lambda$). At this point, it can be argued that some of the  selected channel parameters are very optimistic such as the $d_{TK}$ and $d_{KR}$ distances are chosen small, path loss exponent $\gamma = 2$ is low and a relatively larger keyhole area. This is done only for the sake of computational convenience as we have already provided a theoretical guarantee of the `crossover phenomenon' in the low-SNR regime in Theorem~\ref{eq:theorem2} irrespective of how large the difference is between the path losses $\xi$ and $\varrho$. For these path loss settings, the low-SNR capacity results for the MIMO Rayleigh channel with and without keyhole effect are shown in the Figures~\ref{fig:sevenA},~\ref{fig:sevenB} and~\ref{fig:sevenC}.
\begin{figure*}[!t]
\centering
\begin{subfigure}{.5\linewidth}
\includegraphics[scale = 0.73]{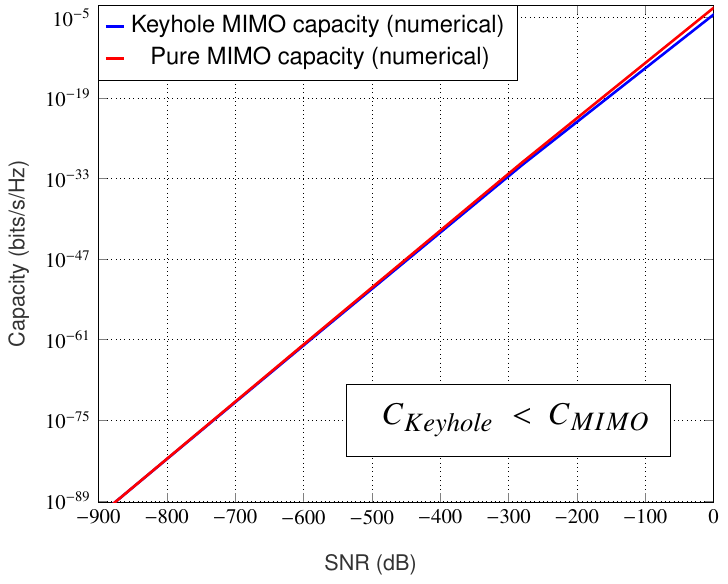}
\vspace{-1.3em}
\caption{}
\label{fig:nineA}
\end{subfigure}%
\begin{subfigure}{.5\linewidth}
  \centering
  \includegraphics[scale = 0.73]{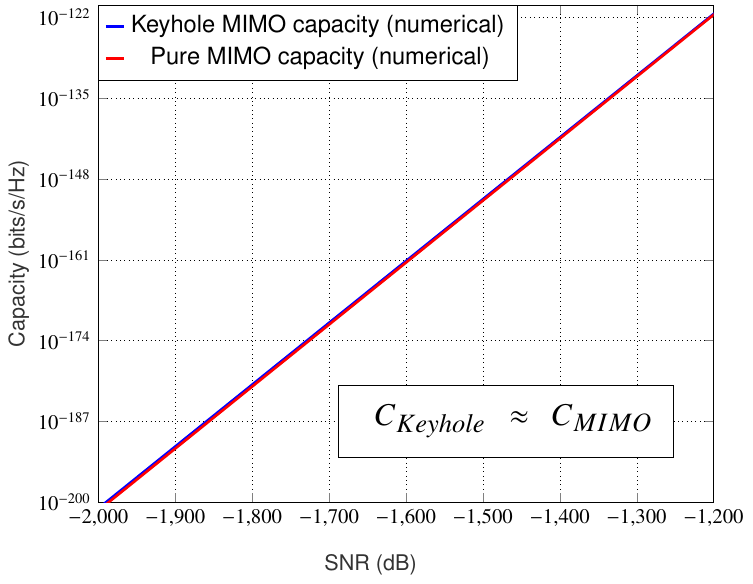}
\vspace{-1.3em}
\caption{}
\label{fig:nineB}
\end{subfigure}
\newline
\begin{subfigure}{1\linewidth}
\vspace{1.0em}
\centering
\includegraphics[scale = 1.45]{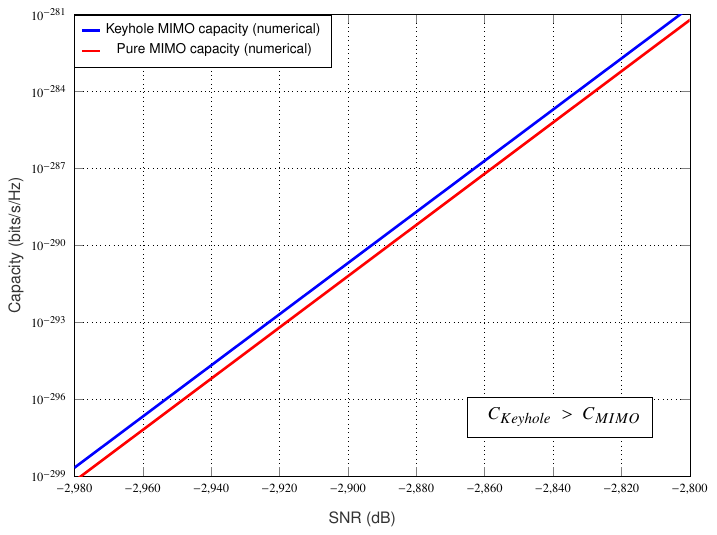}
\vspace{-1.42em}\phantom{xxxxxxxc}\caption{}
\label{fig:nineC}
\end{subfigure}%
\vspace{1.em}
\caption{Exact capacities at low-SNRs for $2 \times 2$ MIMO channel with and without keyhole in IID Nakagami-$m$ fadings ($m = 0.5$) with path loss models~\eqref{eq:powernoKeyhole} and~\eqref{eq:powerKeyhole} included respectively. All path loss related channel parameters remain the same.}
\end{figure*}
\begin{figure*}[!t]
\centering
\begin{subfigure}{.5\linewidth}
\includegraphics[scale = 0.73]{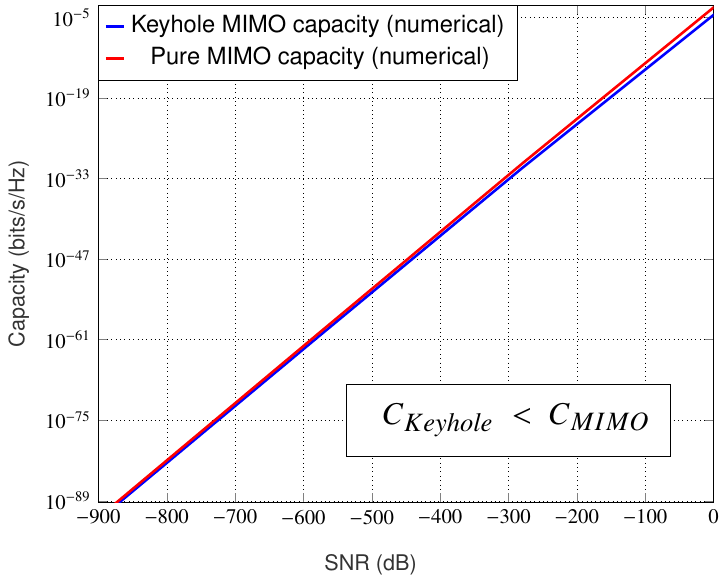}
\vspace{-1.3em}
\caption{}
\label{fig:eightA}
\end{subfigure}%
\begin{subfigure}{.5\linewidth}
  \centering
  \includegraphics[scale = 0.73]{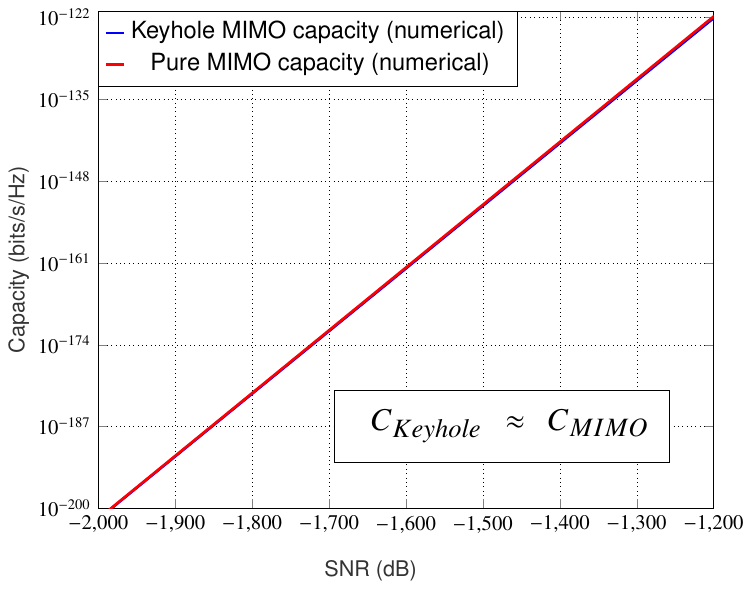}
\vspace{-1.3em}
\caption{}
\label{fig:eightB}
\end{subfigure}
\newline
\begin{subfigure}{1\linewidth}
\vspace{1.0em}
\centering
\includegraphics[scale = 1.45]{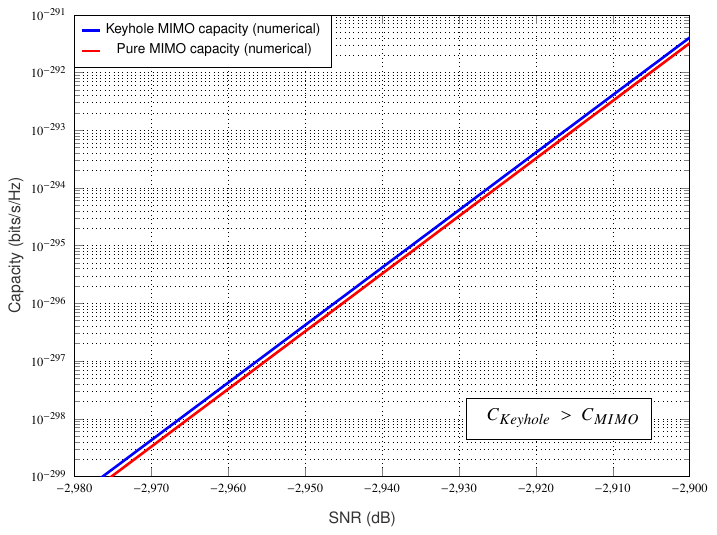}
\vspace{-1.42em}\phantom{xxxxxxxc}\caption{}
\label{fig:eightC}
\end{subfigure}%
\vspace{1.em}
\caption{Exact capacities at low-SNRs for $2 \times 2$ MIMO channel with and without keyhole in IID Nakagami-$m$ fadings ($m = 2$)  with path loss models~\eqref{eq:powernoKeyhole} and~\eqref{eq:powerKeyhole} included respectively. All path loss related channel parameters remain the same except that the distances are $d_{TK} = 6 \lambda$, $d_{KR} = 4 \lambda$ and thus $d_{TR} = 10 \lambda$.}
\end{figure*}
A few remarks are now in order:
\begin{itemize}
\item Comparing the lowerbound in~\eqref{eq:lowerbound:ab} with~\eqref{eq:improve1}, it is clear that the overall effect of incorporating suitable path loss models for the MIMO fading channel with and without keyhole effect, precisely with more severe path loss under keyhole condition than without it, is a delayed crossover point.\\[-0.95em]
\item Allowing path losses to increase in both MIMO fading channels with and without keyhole effect, possibly may be due to increasing distances and/or larger path loss exponents and/or keyhole related impairments such as lower keyhole size, signal attenuation at the keyhole etc., it can be shown that the crossover point will further slide down in the low-SNR regime.\\[-0.95em]
\item The numerical results in the Figures~\ref{fig:sevenA},~\ref{fig:sevenB} and~\ref{fig:sevenC} for the very optimistic large-scale fading conditions assumed there indicates that the crossover point will generally lie in the extremely low-SNR regime. This also suggests that it will be extremely difficult if not impossible to validate this crossover phenomenon via experimental measurements in realistic practical conditions.\\[-0.95em]
\item Though we have utilized distance-dependent path loss models for $\xi$ and $\varrho$ in the build-up to Theorem~\ref{eq:theorem2}, any other suitable large-scale attenuation models (e.g., emperical path loss models) can be applied with no change in the asymptotic low-SNR capacity behaviour suggested by Theorem~\ref{eq:theorem2}.\\[-0.95em]
\end{itemize}

The low-SNR asymptotic capacity improvement of MIMO fading channel under keyhole effect in Theorem~\ref{eq:theorem2} is proved under the assumption of IID Rayleigh fading conditions. Similar low-SNR capacity comparison for the general $r \times t$ MIMO fading channel with and without keyhole effect in IID Nakagami-$m$ fading conditions is slightly constrained due to the lack of eigenmode distributions (joint and marginal) except for some specific cases~\cite{cioffi},~\cite{james}. For the $2 \times 2$ MIMO Nakagami-$m$ fading channel, the joint and marginal eigenmode distribution is derived in a closed form for integer values of $m$~\cite[Theorem~1]{cioffi}, while the joint eigenmode distribution is available for the general $r \times t$ MIMO Nakagami-$m$ fading channel for $m = 0.5$ only~\cite{james},~\cite[p.~107]{anderson}. For these special MIMO Nakagami-$m$ fading channels with path loss conditions kept the same as mentioned earlier (except for $m = 2$ case where we take $d_{TK} = 6 \lambda$, $d_{KR} = 4 \lambda$ and thus $d_{TR} = 10 \lambda$), we have numerically computed the exact MIMO channel capacity with and without keyhole effect in the low-SNR regime, and plotted in Figures~\ref{fig:nineA}-\ref{fig:nineC} and Figures~\ref{fig:eightA}-\ref{fig:eightC}. These numerical evidences and the fact that the rank of MIMO channel has no effect on the capacity at low-SNRs strongly suggests that for the general  $r \times t$ MIMO Nakagami-$m$ fading channel case, the double-faded nature under keyhole effect provides more opportunistic communications at low-SNRs (compared to no keyhole condition), and thus resulting in improved capacity.
\section{Concluding Remarks}\label{sec:conc}
In this paper, we presented simple analytical expressions for the ergodic capacity of the keyhole MIMO channel in IID Nakagami-$m$ fading in the low-SNR regime under the assumption of perfect channel knowledge at both transmitter and receiver. From the asymptotic analysis, we discovered a very surprising or unexpected result that in the low-SNR regime where the spatial degrees of freedom gain has little or no impact, the ergodic capacity of the MIMO fading channel increases when subjected to degenerate keyhole condition, which is in direct contrast of the degrading MIMO capacity behaviour under keyhole effect exhibited in the moderate to high-SNR regime. This low-SNR capacity improvement is due to the double-faded nature of the keyhole MIMO fading channel which increases the possible opportunistic communications created from severe channel fluctuations when compared to the less severe MIMO fading channel without keyhole degeneracy. Further, we have also shown that a simple on-off transmission policy achieves this keyhole channel capacity in the low-SNR regime.
\begin{appendices}
\section{Comparison of Asymptotic Low-SNR capacities\\ derived in terms of the Lambert $W$ function \& Log function in Theorem 2}\label{sec:app:A}
For the ease of exposition, we compare~\eqref{eq:thm_upper_part} and~\eqref{eq:thm_lower_part} keeping only the minimal necessary equivalent expressions as follows:\\[-0.75em]
\begin{itemize}
\item {$\,\,n \, W_{0} (\mathrm{SNR}^{-{1}/{n}} ) \le  \log \left({1}/{\mathrm{SNR}}\right)\,\,$ for $\,\,n > 0$}:\\[0.25em]
For $x >> 1$: $y=xe^x \Leftrightarrow x=W_0 (y)$. Applying the log function on both sides of the last equality gives:
\begin{align}\label{eq:inequality1_1}
\log(y)&= x + \log(x) \nonumber \\[0.35em]
       &= W_0 (y) + \log(x) \nonumber\\[0.35em]
& \ge W_0 (y)
\end{align}
For $\mathrm{SNR} \rightarrow 0$ and any $n > 0$, the $y=\textrm{SNR}^{-{1}/{n}}$ substitution in~\eqref{eq:inequality1_1} is valid, and gives
$
\log(\mathrm{SNR}^{-{1}/{n}}) \ge W_0 (\mathrm{SNR}^{-{1}/{n}})
$
which proves the inequality.\\[-0.25em]
\item {$\,\,|\,n W_{-1} (-\mathrm{SNR}^{-{1}/{n}})\,| \ge  \Big|\log \left(\frac{1}{\mathrm{SNR}}\right)\Big|\,\,$ for $\,\,n < 0$}:\\[0.25em]
For $x<<-1$, we note that $y=x e^x \Leftrightarrow x=W_{-1}(y)$ and ${-1}/{e}<y<0$. Consider $-y=-x e^x$ and apply the log function on the both sides:
\begin{align}\label{eq:inequality1_2}
\log(-y)&= x + \log(-x) \nonumber \\[0.35em]
        &= W_{-1}(y) + \log(-x) \nonumber \\[0.35em]
\Rightarrow  |\log(-y)| &\le |W_{-1}(y)|
\end{align}
where the last inequality is due to the facts that $W_{-1}(y)<<-1$, $\log(-x)>>0$ and $\log(-y)<<0$. With
the valid $y=-\mathrm{SNR}^{-{1}/{n}}$ substitution in~\eqref{eq:inequality1_2} where $\mathrm{SNR} \rightarrow 0$ and $n < 0$, we get
$
|W_{-1}(-\textrm{SNR}^{-{1}/{n}})| \ge |\log(\mathrm{SNR}^{-{1}/{n}})\,|
$
which proves the inequality.
\end{itemize}

\section{On-Off Power Control is asymptotically Optimal\\
for the Keyhole MIMO Channel at Low-SNR}\label{sec:app:B}
The on-off power $P ({\lambda})$ equals $P_0$ for $\lambda > \lambda_0$ and zero otherwise; the constant $P_0$ satisfies $\E [ P ({\lambda})] = \mathrm{SNR}$. Thus,
\begin{align}\label{eq:def_on_off}
P(\lambda) = \begin{cases}
\dfrac{\mathrm{SNR}}{\mathrm{Prob}({\lambda} >  \lambda_0)}\,\,\,\,\,\,\mathrm{if}\,\,\,\lambda > \lambda_0,\\[.75em]
        \phantom{x}0\,\,\,\,\,\phantom{xxxxxxxxx}\mathrm{otherwise}.
\end{cases}
\end{align}
where the channel threshold $\lambda_0$ is chosen same as the cut-off obtained for the optimal waterfilling power allocation.

The ergodic rate achievable with this on-off transmission scheme is
\begin{align}
R  \,&=\,  \int_{\lambda_0}^{\infty} \log(1 +  \lambda P_0) f_{\bm{\lambda}} (\lambda) d\lambda  \\
                  &\geq \, \log(1 +  \lambda_0 P_0) \int_{\lambda_0}^{\infty} f_{\bm{\lambda}} (\lambda) d\lambda \\
                  &= \, \log\left(1 +  \dfrac{\lambda_0 \, \mathrm{SNR}}{\mathrm{Prob}({\lambda} >  \lambda_0)}\right) \mathrm{Prob}(\bm{\lambda} >  \lambda_0). \label{eq:onoffsum:def}
\end{align}
With the low-SNR approximation in~\eqref{eq:bessel_approx} applied to~\eqref{eq:cap:csit0}, the tail probability $\mathrm{Prob}({\lambda} >  \lambda_0)$ is obtained as
\begin{align}\label{eq:tempey1}
\mathrm{Prob}({\lambda} >  \lambda_0)  \approx \dfrac{\sqrt{\pi}}{\Gamma(c_t)\Gamma(c_r)} I_1 \left(\frac{\lambda_0}{b_t b_r}\right)
\end{align}
where, in turn, $I_1 (\cdot)$, as defined in~\eqref{eq:I_definitions}, is approximated with the first-term
only in~\eqref{eq:gamma_approx} (valid for low-SNR conditions) to further simplify~\eqref{eq:tempey1} as
\begin{align}\label{eq:tempey2}
\mathrm{Prob}({\lambda} >  \lambda_0) \approx \dfrac{\sqrt{\pi}}{\Gamma(c_t)\Gamma(c_r)} e^{-2\sqrt{\frac {{\lambda}_0}{b_t b_r}}} \left(\dfrac {{\lambda}_0}{b_t b_r}\right)^{\frac{c_t + c_r}{2} - \frac{3}{4}} \cdot
\end{align}
Using~\eqref{eq:cutoff131},~\eqref{eq:tempey2} and recalling $\mu_0:= \dfrac{\lambda_0}{b_t b_r}$, we get
\begin{align}\label{eq:log_simplified}
\dfrac{\lambda_0 \, \mathrm{SNR}}{\mathrm{Prob}({\lambda} >  \lambda_0)} \approx \left(\dfrac {{\lambda}_0}{b_t b_r}\right)^{-\frac{1}{2}},
\end{align}
which approaches to zero as $\lambda_0$ goes to infinity (at low-SNR). Combining~\eqref{eq:log_simplified} and~\eqref{eq:onoffsum:def} with the
$\log(1 + x) \approx x$ approximation, we conclude that
\begin{align}\label{eq:onoff_bound}
R  \,\geq\, \lambda_0 \, \mathrm{SNR},
\end{align}
\noindent
where the lower bound in~\eqref{eq:onoff_bound} above is the asymptotic low-SNR capacity $C$. This guarantees that the on-off power scheme is capacity-achieving in the asymptotic regime of low-SNR. Notice that the on-off scheme requires only 1-bit CSI-T feedback (i.e., good or bad channel state). This is practically attractive in low-SNR conditions as binary CSI-T feedback can be made more reliable than high-resolution CSI-T feedback for a given fixed amount of resources reserved for feedback transmissions.

\begin{figure}[H]
\centering
\includegraphics[scale = 1.25]{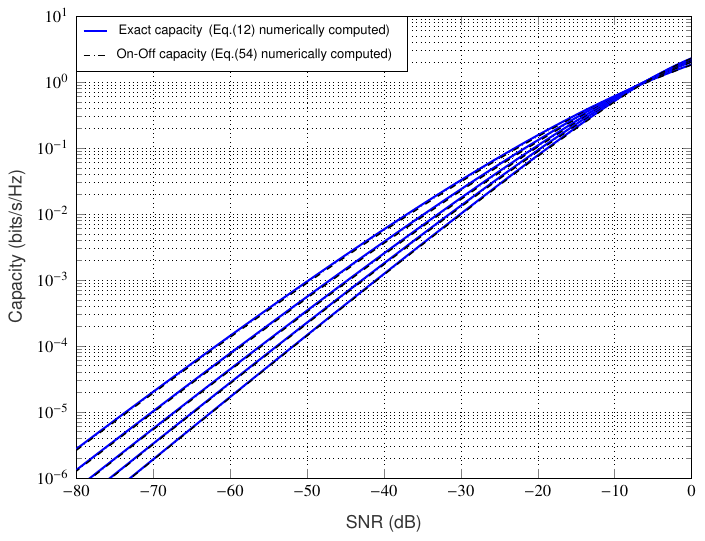}
\vspace{1em}
\caption{Ergodic rates of $2 \times 2$ keyhole MIMO channel in Nakagami-$m$ fading with CSIT for On-off \& Waterfilling power schemes at low SNRs. For simplicity, we keep $m_r = m_t$ and $\Omega_r = \Omega_t = 1$. For each power scheme (On-off or Waterfilling), the curves correspond (from the left side), in descending order, to $m_r = m_t = \sfrac{1}{2}, 1, 2, 4, 10$.}
\label{fig:seven}
\end{figure}

We observe from Fig.~\ref{fig:seven} that the on-off rates are almost indistinguishable from the exact capacity curves for fading conditions varying from severe ($m = 0.5$) to moderate ($m = 2$) and finally to mild ($m = 10$) levels, while the SNR varies from moderately low to extremely low values. That is, the simple on-off transmission strategy achieves near-optimal performance at realistic low SNRs for a wide range of practical fading scenarios. As already mentioned in the previous subsection, the low-SNR capacity improvement with decreasing shape parameter $m$ is due to the fact that frequent channel fluctuations (e.g., peaks and nulls) knowledge at the transmitter side in the low-SNR regime is hugely beneficial to dynamically allocating power and transmitting opportunistically only when the channel realization is in good condition.


%
%
%
%
\end{appendices}

%

\end{document}